\pdfoutput=1
\documentclass[nopubldata]{lpor2014}
\usepackage{blindtext}
\usepackage{hyperref}
\usepackage{amsmath}
\usepackage{cancel}
\usepackage{eqnarray}
\usepackage{soul}
\hypersetup{%
  colorlinks,
  plainpages=false,
  pdfpagelabels,
  breaklinks,
  pdfview=Fit,
  bookmarksopen,
  bookmarksnumbered,
  linkcolor=black,
  anchorcolor=black,
  citecolor=black,
  filecolor=black,
  urlcolor=LPRred
  }%
\graphicspath{{./figs/}}
\setpages[]{}
\setvolume[0]{0}
\setyear{2011}%
\setdoi{201100000}%
\oddheadlogotext{}{Guide}

\newcommand*\samethanks[1][\value{footnote}]{\footnotemark[#1]}

\begin{document}

\abstract{%
 Short laser pulses are widely used for controlling molecular rotational degrees of freedom and inducing molecular alignment, orientation, unidirectional rotation and other types of coherent rotational motion. To follow the ultra-fast rotational dynamics in Òreal timeÓ, several techniques for producing Òmolecular moviesÓ have been proposed based on the Coulomb explosion of rotating molecules, or recovering molecular orientation from the angular distribution of high-harmonics. The present work offers and demonstrates a novel non-destructive optical method for direct visualization and recording of ÒmoviesÓ of coherent rotational dynamics in a molecular gas. The technique is based on imaging the time-dependent polarization dynamics of a probe light propagating through a gas of coherently rotating molecules. The probe pulse continues through a radial polarizer, and is then recorded by a camera. We illustrate the technique by implementing it with two examples of time-resolved rotational dynamics: alignment-antialignment cycles in a molecular gas excited by a single linearly polarized laser pulse, and unidirectional molecular rotation induced by a pulse with twisted polarization. This method may open new avenues in studies on fast chemical transformation phenomena and ultrafast molecular dynamics caused by strong laser fields of various complexities.   }

\title{Optical imaging of  coherent molecular rotors}

\author{J\'er\'emy Bert\inst{1}\thanks{These authors contributed equally to this work.}, Emilien Prost\inst{1}\samethanks, Ilia Tutunnikov\inst{2}\samethanks,  Pierre B\'ejot\inst{1}, Edouard Hertz\inst{1}, Franck Billard\inst{1}, Bruno Lavorel\inst{1}, Uri Steinitz\inst{2,3}, Ilya Sh. Averbukh\inst{2,*}, and Olivier Faucher\inst{1,*}}%
\authorrunning{J. Bert et al.}
\mail{\email{olivier.faucher@u-bourgogne.fr, ilya.averbukh@weizmann.ac.il}}

\institute{%
Laboratoire Interdisciplinaire CARNOT de Bourgogne, UMR 6303 CNRS-Universit\'e de Bourgogne, BP 47870, 21078 Dijon, France;
\and
AMOS and Department of Chemical and Biological Physics, The Weizmann Institute of Science, Rehovot 7610001, Israel;
\and 
Soreq Nuclear Research Centre, Yavne, Israel}  


\begin{acknowledgement}
This work was supported by  the CNRS, the ERDF Operational Programme - Burgundy,  the EIPHI Graduate School (Contract No. ANR-17-EURE-0002), the Associate (CNRS\&Weizmann)  International ImagiNano Laboratory, and the Israel Science Foundation (Grant No. 746/15). Calculations were performed using HPC resources from DNUM-CCUB (Universit\'e de Bourgogne).  I.A. acknowledges support as the Patricia Elman Bildner Professorial Chair. This research was made possible in part by the historic generosity of the Harold Perlman Family.
\end{acknowledgement}

\maketitle

\section{Introduction}
\label{sec:intro}
Resolving the motion of molecular nuclei during chemical transformations, or when under the influence of ultrafast external fields, is of extreme importance in Chemistry and Physics \cite{Zewail2009,Dwayne2017}. The ultimate goal is recording the so-called ``molecular movies'' unveiling the corresponding molecular dynamics in ``real time''.  Filming molecules involves many different time scales, from short ones relevant to the electronic dynamics, to the longest ones corresponding to molecular rotation.  Recent advances in high-intensity electron and X-ray pulsed sources \cite{Guhr2016,Young_2018}  made it possible to directly observe atomic motions and electron dynamics happening on intramolecular scale. As it is easier to follow and control the slow motion, molecular rotational dynamics became a testing ground for developing new methods for molecular ``cinematography'', and various approaches for quantum control of molecular dynamics. In particular, detailed ``movies'' were recorded for imaging unidirectional  rotation of linear molecules subject to strong laser pulses with crossed linear polarizations \cite{Lin92_2015,Mizusee1400185}, and extremely rich rotational dynamics of molecules undergoing field-free alignment-antialignment cycles \cite{Karamatskos2019}.
Detection methods for probing the angular localization of molecules resulting from strong laser field interactions can be divided into two categories. The first  one is based on fragmentation of the molecules, while the second one relies on optical detection. In the first category, short and intense resonant laser pulses are used for breaking molecular bonds through a resonant dissociation process followed by ionization of the fragments \cite{Larsen17_1999} or through  Coulomb explosion \cite{Litvinyuk90_2003}. In  the case of a linear molecule, molecular fragments are assumed to recoil along the line defined by the molecular axis at the moment of molecular ionization. This allows retrieving the orientation of the molecule from velocity map imaging  (VMI) \cite{Helm70_1993} of the photofragments. Because the technique is sensitive to the charge trajectories projected on the plane of a detector, the  VMI  provides a two-dimensional information about the angular distribution of the molecules \cite{Stapelfeldt26_2003}. In contrast,  a full 3D information  can be accessed  by  cold target recoil ion momentum spectroscopy (COLTRIMS)  \cite{Dorner330_2000}, where both electron and ion momenta  are imaged in coincidence \cite{Lin92_2015}. Methods based on charge momentum imaging provide a wealth of information about the molecular dynamics and therefore constitute powerful tools for their full characterization. However, such techniques require sophisticated apparatus, long acquisition times and meticulous data post processing, and are applicable only to the rarefied and cold molecular gases.   In contrast, optical techniques do not require highly dedicated apparatus,  can be easily implemented, and are   compatible with a wide range of gas pressures and temperatures \cite{Faucher100_2011}. So far, optical detection schemes have been limited to one-dimensional measurements of the ensemble averaged quantities, like orientation factor, $\left<\cos\theta\right>$ \cite{Fleischer16_2011} and alignment factor, $\left<\cos^2\theta\right>$\cite{Renard15_2003},  where $\theta$ is the Euler angle between the molecule-fixed and a space-fixed axes. They have also been  used to measure higher-order moments $\left<\cos^n\theta\right>$, with $n>2$, of the  angular distribution by using high-harmonic generation \cite{Marangos26_2013}. Recently, angle-resolved high-order-harmonic spectroscopy was used for generation of a molecular ``rotational movie'' with the help of iterative machine learning procedure \cite{He2019}.

The present work introduces a new optical technique for direct time-resolved imaging of molecular rotation. We demonstrate a direct visualization of the angular localization of molecules in space.  The technique is applied here to  observing ultrafast temporal  alternation between molecular alignment and antialignment in a molecular gas excited by a short non-resonant linearly polarized pulse, as well as to monitoring coherent  unidirectional  rotation of molecules in the case of excitation by a pulse with twisted polarization.

 \begin{figure}
  \includegraphics*[width=\linewidth]{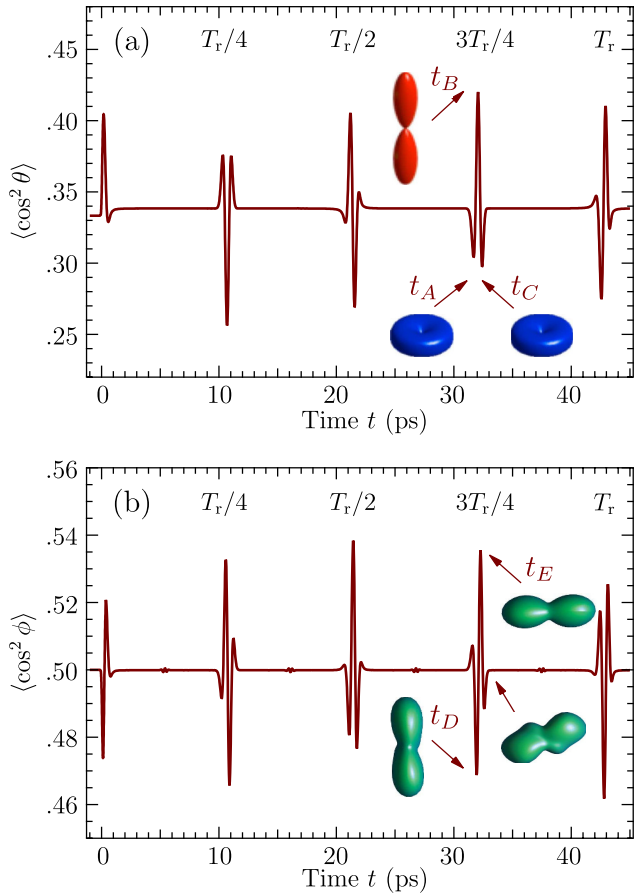}
  \caption{a)  Temporal evolution of the alignment factor $\left<\cos^2\theta\right>$ of CO$_2$ molecules aligned by a  linearly-polarized laser pulse (centered at $t=0$). The angle   $\theta$ defines the orientation of the molecular axis with respect to the direction of the laser field. The angular distributions for aligned ($t_{B}=32.06$ ps)  and antialigned ($t_{A}=31.68$ ps and $t_{C}=32.4$ ps)   molecules are depicted in red and blue,    respectively.   b)  Expectation value $\left<\cos^2\phi\right>$ describing the alignment  of the molecules  in the plane of rotation  of a driving field exhibiting a twisted linear polarization.  The angular distributions of the molecular axes at time  $t_{D}=31.94$ ps,  $t_{E}=32.27$ ps, and in between  are represented in green.}
  \label{fig-Align}
\end{figure}

\section{Concept}
\label{sec:concept}

\subsubsection{Aligned molecules}
\label{sssec:alignment}  
Laser-induced  molecular alignment results from the anisotropic interaction  between the electric field   of a strong laser pulse  and the induced molecular dipole. The concept was introduced in the early 90s \cite{Normand20_1992,Friedrich23_95} and was initially  investigated in the adiabatic regime, using pulse durations much longer that the rotational period  $T_{\textrm{r}}$ of the molecule.
In general, long pulses allow achieving high  degree of alignment, even at moderate temperatures. However, a drawback of the adiabatic interaction is that the alignment exists only in the presence of a strong  external field,  which may complicate the interpretation of the experimental results and also limits the potential applications. As an alternative, impulsive alignment was proposed \cite{Seideman83_1999,Ortigoso110_1999} and developed, allowing preparation of aligned molecular samples under field-free conditions \cite{Vrakking87_2001}. Although our imaging technique is in principle applicable to both the adiabatic and  non-adiabatic cases, the present work  focuses  on the latter regime, that represents  a widely used  common approach for aligning molecules with lasers (more details  about laser-induced molecular alignment can be found in Refs. \cite{Stapelfeldt75_2003,Ohshima29_2010,Fleischer52_2012}).

 Figure \ref{fig-Align}a  depicts a calculated alignment factor of CO$_2$  molecules at room temperature kicked by a nonresonant short laser pulse linearly polarized along $z$ axis. 
  The alignment process is studied  by  solving the Liouville-von Neumann equation for molecules driven by a Gaussian laser pulse with  peak intensity of 20 TW/cm$^2$ and duration (FWHM) of 100 fs, which is applied at time $t=0$. The theoretical model, described  in Ref. \cite{Prost96_2017}, allows  computing molecular  rotational dynamics   for any polarization of the excitation pulse.  The values of the alignment factor $\left<\cos^2\theta\right>>1/3$ signify the  preferential alignment of the molecular axis along the polarization direction of the field. A state in which the molecules are preferentially localized in a plane perpendicular to the field is called antialignment, and it is characterized by $\left<\cos^2\theta\right><1/3$. During the field-free evolution of the quantum system after the turnoff of the aligning field ($t> 0.1$  ps), the anisotropic angular molecular distribution  alternates between alignment and antialignment within each \textit{rotational revival} recurring with a period given by $T_{\textrm{r}}/4$ (see, e.g. the corresponding  angular distributions of  molecules depicted around the third fractional revival at $3T_{\textrm{r}}/4$). Between the revivals, a residual permanent alignment is manifested by a value of $\left<\cos^2\theta\right>$ standing slightly above its isotropic value of 1/3 \cite{Poulsen121_2004}.

\begin{figure}
  \includegraphics*[width=\linewidth]{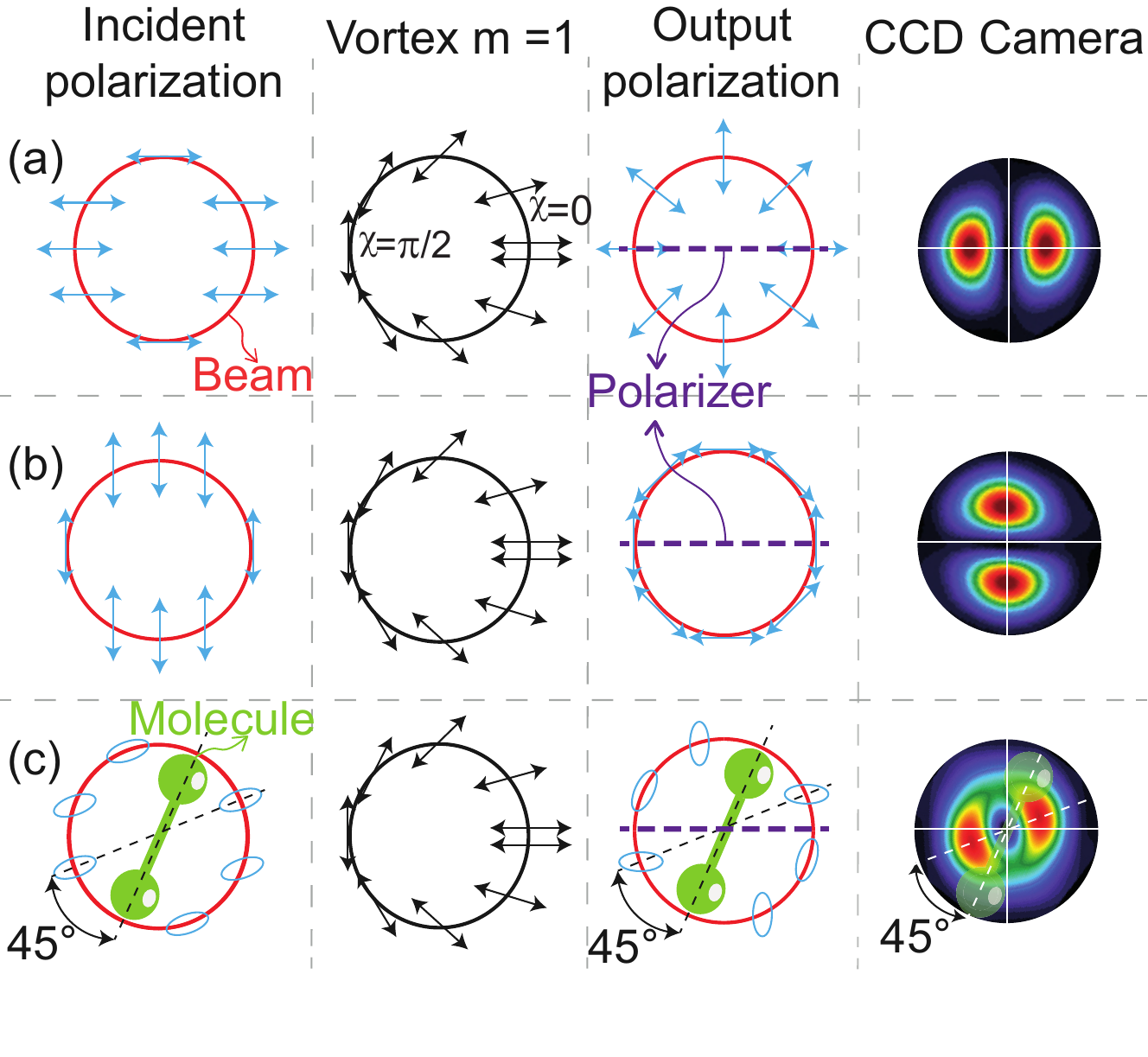}
  \caption{A sketch depicting the conversion of polarization of the incident light beam (left column)  into an intensity pattern (right column). The resulting intensity patterns are captured by a CCD camera after the light passes through a $m=1$ vortex plate (second column) and a linear polarizer (third column, dashed line). The $m=1$ vortex plate is illustrated by arrows oriented along the fast-axes of the half-wave retarders (black arrows).  In a) and b), the horizontal and vertical polarization of the incident linearly polarized light (blue arrows) is converted into intensity patterns oriented along the horizontal and vertical directions, respectively. In c), the incident elliptically polarized light (blue ellipses) is converted into an intensity pattern whose symmetry axis is parallel to the major axis of the polarization ellipse (and at -45\textDegree to the slow axis of the medium).}
  \label{fig-vortex}
\end{figure}
\subsubsection{Spinning molecules }
\label{sssec:spinning}
If instead of exciting the molecules by a linearly polarized laser pulse, a field with twisted linear polarization \cite{Karras114_2015}  is applied to the system, then a unidirectional rotation of the molecules can be induced in the plane of polarization, according to the mechanism described in Refs. \cite{Fleischer11_2009,Kitano103_2009}. The molecular angular momentum gets oriented along the direction of the field propagation, and the sense of rotation can be controlled by adjusting the phase of the shaped pulses \cite{Prost96_2017}.  Other field configurations leading to similar effect include a chiral trains of pulses \cite{Zhdanovich2011}, a pair of delayed polarization-crossed pulses \cite{Fleischer11_2009,Khodorkovsky83_2011,Gershnabel120_2018}, and an optical centrifuge \cite{Yuan26042011,Korobenko2014,Karczmarek82_1999}.

Field-free evolution of the molecular ensemble excited by a  twisted field can be captured by measuring the value of  $\left<\cos^2\phi\right>$, where $\phi$ is the azimuthal Euler angle (relative to the $x$ axis) in the $xy$ plane in which the polarization of the field is confined. This observable allows to estimate when molecules are directed along the $x$ axis ($\left<\cos^2\phi\right>>1/2$), or the $y$ axis ($\left<\cos^2\phi\right><1/2$), while the isotropic distribution corresponds to $\left<\cos^2\phi\right>=1/2$. Figure \ref{fig-Align}b shows the simulated result  for CO$_2$ after the molecules have been excited by a  sequence of two orthogonally polarized pulses of same duration (100 fs) and peak intensity (20 TW/cm$^2$) separated by 145 fs and phase synchronized. One can see that during  the revivals, the molecular behavior resembles  the rotation of an aircraft propeller. This fast coherent molecular rotation has been  detected in the past with the help of  rotational Doppler effect resulting from the exchange of energy and angular momentum between the spinning molecules and a circularly-polarized probe optical field \cite{Korech7_2013,Faucher94_2016}.

\section{Experimental results}
\label{sec:experiment}
\subsection{Experimental method}
\label{ssec:method}
\begin{figure*}
  \includegraphics*[width=\linewidth]{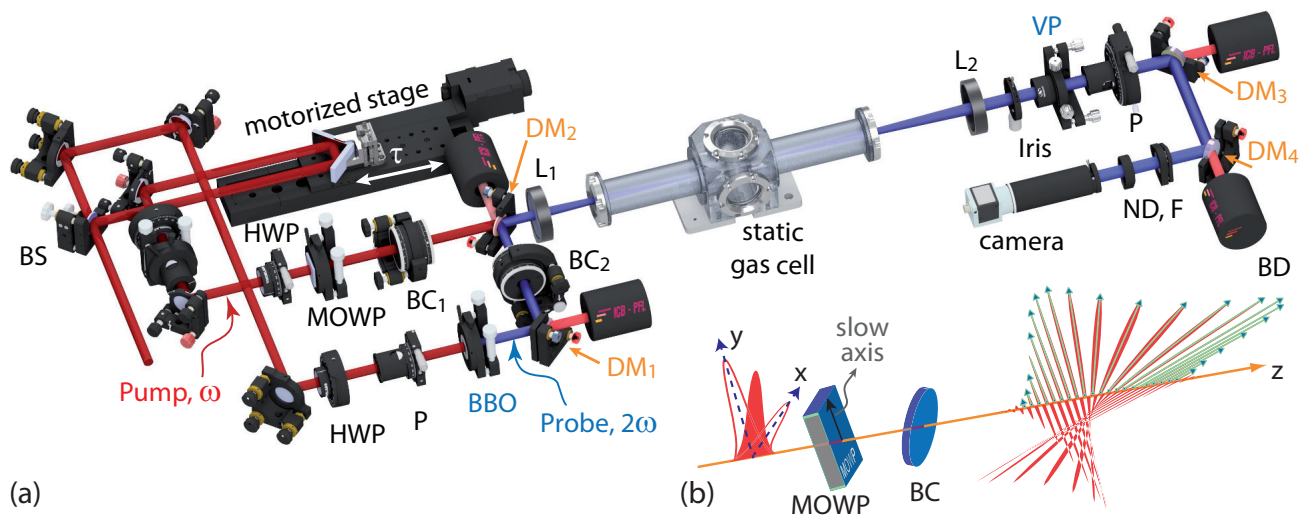}
  \caption{ a) Experimental setup for producing and imaging coherent molecular rotation. The left side of the figure is related to the preparation of a coherent excitation of  gas phase molecules  by a  polarization shaped femtosecond laser pulse (pump).   The right side describes the  imaging device used to record the spatial orientation of the molecules read by a probe pulse.   (BS, beam splitter; HWP, half-wave plate; P, polarizer; BBO, type-I phase-matching $\beta$-barium borate crystal; MOWP,  multiple-order wave plate; BC, Berek compensator; DM, dichroic mirror; L, lens; VP, vortex plate; F, UV bandpass filter; BD, beam dump; ND, neutral density; F, filter. b)  The polarization shaper consisting in  a MOWP and  a BC generates  two partially  time overlapped,  cross-polarized in phase (or out of phase) short laser pulses leading to a field with  a twisted linear polarization  \cite{Karras114_2015}. }
  \label{fig-setup}
\end{figure*}
From the point of view of optics, a linear molecule can be seen as a birefringent wave plate having a slow principal axis oriented along the molecular bond. The refractive index for a light polarized along the molecule axis therefore differs from that for a light polarized along the perpendicular direction. Here, we consider an optical imaging of a gas of linear molecules that has been transiently aligned or spun by a non-resonant laser pulse (pump). To this end, the sample is probed by a time delayed circularly-polarized short pulse (probe). During the probe propagation the medium is assumed to be stationary. The probe field is decomposed along the two principal axes of the anisotropic medium. Each polarization component propagates through the medium under a different refractive index resulting in accumulation of relative phase between them. After passing the medium, the probe becomes elliptically polarized with the major axis of the ellipse oriented at $\pm \pi/4$ to the molecular axis (the sign depends on the sense of circular polarization).  In order to locate the orientation of the ellipse, the probe field is directed to a two-dimensional (2D) polarization analyzer  combining a vortex plate and a  linear polarizer. The zero-order vortex $m=1$  retarder consists of half-wave plates whose  local orientations  $\chi$ vary  continuously with the azimuthal angle $\Phi$ of the plate  according to the law $\chi=m\Phi/2$ (see Figure \ref{fig-vortex}).

The principle of action of the 2D polarization analyzer is presented in Figure \ref{fig-vortex}. As depicted in Figures \ref{fig-vortex}a and \ref{fig-vortex}b the vortex plate converts a linear horizontal polarization in to a radial polarization, and inversely, a vertical linear polarization into an azimuthal polarization. The direction of polarization of the incident light is retrieved by imaging the intensity of the transmitted beam on a charge-coupled device (CCD) camera. To this aim, a linear polarizer oriented parallel to the symmetry axis of the vortex plate is inserted between the vortex plate and the CCD camera. As shown, in both cases the symmetry axis of the intensity pattern rendered on the CCD camera is parallel to the major axis of the polarization ellipse of the incident light.

 When the  molecules are randomly oriented, the circular polarization of the probe  passing through the vortex plate results in a balanced superposition of radial and azimutal polarization producing an isotropic intensity distribution on the CCD camera. However, when the isotropy of the medium is broken, i.e. during a molecular revival,  the  probe light  becomes elliptically polarized. This situation is  illustrated in Figure \ref{fig-vortex}c, where we assume that  the major axis of the  ellipse before the vortex plate is oriented at $- \pi/4$ with respect to the molecule axis. After the vortex plate, all ellipses are differently tilted depending on the position across the beam profile,except at $-\pi/4$ ($\pi/4$) where the major axes of the  ellipses are parallel (perpendicular) to the linear polarizer. The resulting spatial distributions of elliptical polarization after the polarizer leads to an intensity pattern reflecting the symmetry of the polarization ellipse. A sequence of images taken during the revival event allows to record the motion of the molecule in the plane of polarization of the probe.

\subsection{Experimental setup}
\label{ssec:setup}
The experiment is based on a femtosecond Ti:sapphire chirped-pulse amplifier producing pulses of 100 fs duration with a repetition rate of 1 kHz and a maximum energy of 3 mJ. The  overall experimental setup  is shown in Figure \ref{fig-setup}a. The polarization of the pump pulse  ($\omega$)  is shaped  by   a multiple-order wave plate (MOWP) combined with a Berek compensator (BC1). By changing the polarization direction of the incident field it is possible to generate at the exit of the shaper \cite{Karras114_2015} a pulse  exhibiting a fixed  linear  polarization or  a twisted linear polarization as illustrated in Figure \ref{fig-setup}b. The former is used to  impulsively align the molecules, whereas the latter produces transiently spinning  molecules, as  previously described in Sec. \ref{sec:concept}. The probe beam is frequency doubled (2$\omega$) with a BBO crystal and then  overlapped with the pump beam. The Berek compensator (BC2), placed  before the dichroic mirror (DM2), has its  retardance adjusted in order to ensure that the probe pulse is  circularly  polarized after reflection by DM2. A motorized retroreflector is introduced in the optical path of the pump beam for controlling the optical delay $\tau$ between the pump and probe beam.  Both beams are focused with a lens (L1) in a static cell filled with  room temperature CO$_2$ gas. A second lens (L2) is inserted before the vortex plate in order to collimate the  probe beam. To avoid the pump light reaching the detector, two dichroic mirrors (DM) and a filter (F) are placed before the CCD camera.

\subsection{Results and discussions}
\label{ssec:results}
\begin{figure}
  \includegraphics*[width=\linewidth]{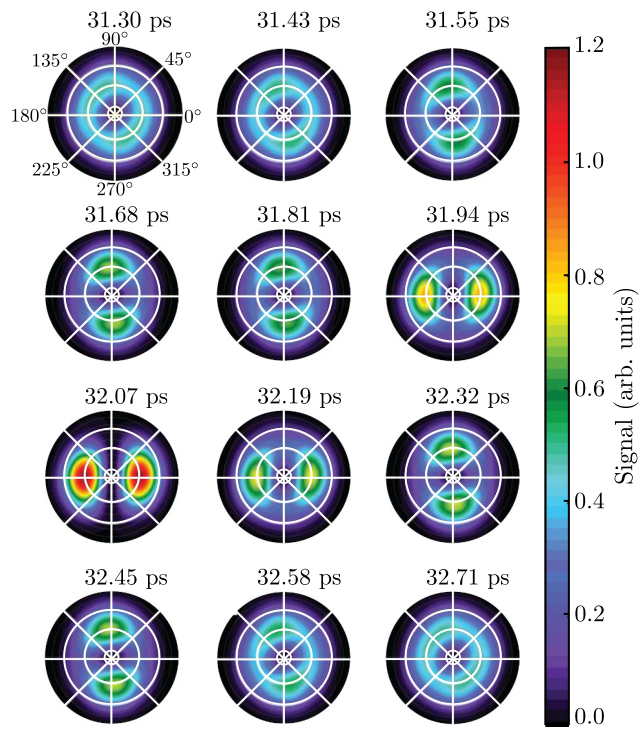}
  \caption{ Optical imaging of  aligned molecules. Snapshot captured  for different pump-probe delays during the $3T_{\textrm{r}}/4$ revival of CO$_2$.  The delays $\tau= 31.68$, $32.07$, and $32.45$ ps correspond to the three events  $t_A$  (antialignment), $t_B$  (alignment), and $t_C$  (antialignment)  defined in Figure \ref{fig-Align}a, respectively. The pump pulse is horizontally polarized.}
  \label{fig-Align3tr/4}
\end{figure}
We  start with   the results related to  molecular alignment imaging.  The snapshots presented in Figure \ref{fig-Align3tr/4} have been recorded in 1 bar of CO$_2$  when the pump-probe delay is tuned  over the third alignment revival $3T_{\textrm{r}}/4$  discussed in Figure \ref{fig-Align}a. The   movie of the whole sequence can be found in the Supporting Information. The aligning pulse (pump) is horizontally polarized in the plane of the camera. Recall that depending on the handedness of the probe's polarization the present technique produces a $\pm\pi/4$ angle between the molecules and the major axis of the polarization ellipse of the probe field imaged on the detector. Therefore, for convenience, all images presented in this work have been rotated by $\mp\pi/4$, so that the symmetry axis of the images coincides with the molecular axis.

Figure \ref{fig-Align3tr/4} shows two types of images, one with an elongated intensity distribution  along the horizontal axis, i.e. parallel  to the pump polarization, and  another one elongated along the vertical axis.  Note that the  vortex plate  transforms the incident Gaussian probe beam TEM$_{00}$ profil into a Laguerre-Gaussian beam LG$_{01}$, which explains  the   attenuation of the signal observed near the singularity at the center of each snapshot.  In the first group of images, the molecules are aligned along the polarization direction of the pump laser field, whereas in the second group they are delocalized (antialigned) in the plane perpendicular to the pump. The maximal degree of alignment is observed around $\tau=32.07$ ps, which is in good agreement with the simulation  presented in Figure \ref{fig-Align}a,  predicting the moment of maximal alignment around $t_B=32.06$ ps for the  $3T_{\textrm{r}}/4$ revival. The antialignment reaches its maximum value for $\tau=31.68$  and $32.45$ ps, which is also consistent  with  the calculated values of $t_A$ and $t_C$. The results of Figure \ref{fig-Align3tr/4} provide a direct visualization of the complex spatio-temporal dynamics of the molecular alignment process. In the course of each revival, molecules  toggle between alignment and antialignment; two different coherent states of the   molecular rotor.

\begin{figure}
  \includegraphics*[width=\linewidth]{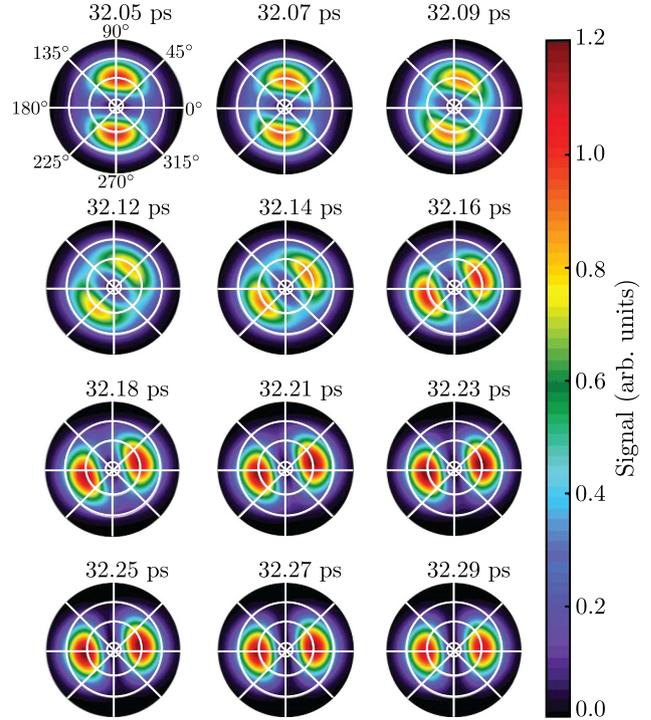}
  \caption{Optical imaging of spinning molecules. The images captured by the camera for the  delays $\tau=$ 32.05 and 32.29 ps correspond to the  events  $t_D$ and $t_E$ defined in Figure \ref{fig-Align}b, respectively. The polarization of the driving  pulse describes a clockwise rotation in the image plane.}
  \label{fig-udr}
\end{figure}
The images presented in Figure \ref{fig-udr} have been recorded when the molecules are spun by a pulse exhibiting a twisted linear polarization. The latter executes a clockwise rotation corresponding to a quarter turn, starting from the vertical direction and ending horizontally. It triggers a unidirectional rotation of the molecules, which is here captured during the third revival of CO$_2$. Note that  similar dynamics can be observed for other revivals. The accuracy of the imaging technique can be attested by comparing the measurements to the simulations of $\left<\cos^2\phi\right>$. In particular, one can observe the images recorded  around the specific times $t_D$ and $t_E$ introduced in  Figure \ref{fig-Align}b. The local minimum $t_D$ of the function $\left<\cos^2\phi\right>$ implies a preferential alignment of the molecules along the $y$ axis, which is well observed if one inspects the images recorded at the corresponding time $\tau= 32.05$. In contrast, the local maximum at $\tau=t_E$ suggests that the molecular axes are directed along the $x$ axis, which is also confirmed by the images obtained at $\tau= 32.29$ ps. The coherent rotation of the molecules, which, it should be noted, can not be inferred from  $\left<\cos^2\phi\right>$, is noticeable when one observes the successive  images recorded between $\tau=32.05$ and 32.29 ps. The clockwise sense of rotation is also clearly evidenced by the corresponding movie provided in the Supporting Information. As shown in the next section, from the previous image sequence it is possible to extract quantitative information about the orientation and speed of the molecules.

\section{Data analysis and numerical simulations}
\label{sec:model}
\label{sec:model}
\begin{figure}
  \includegraphics*[width=\linewidth]{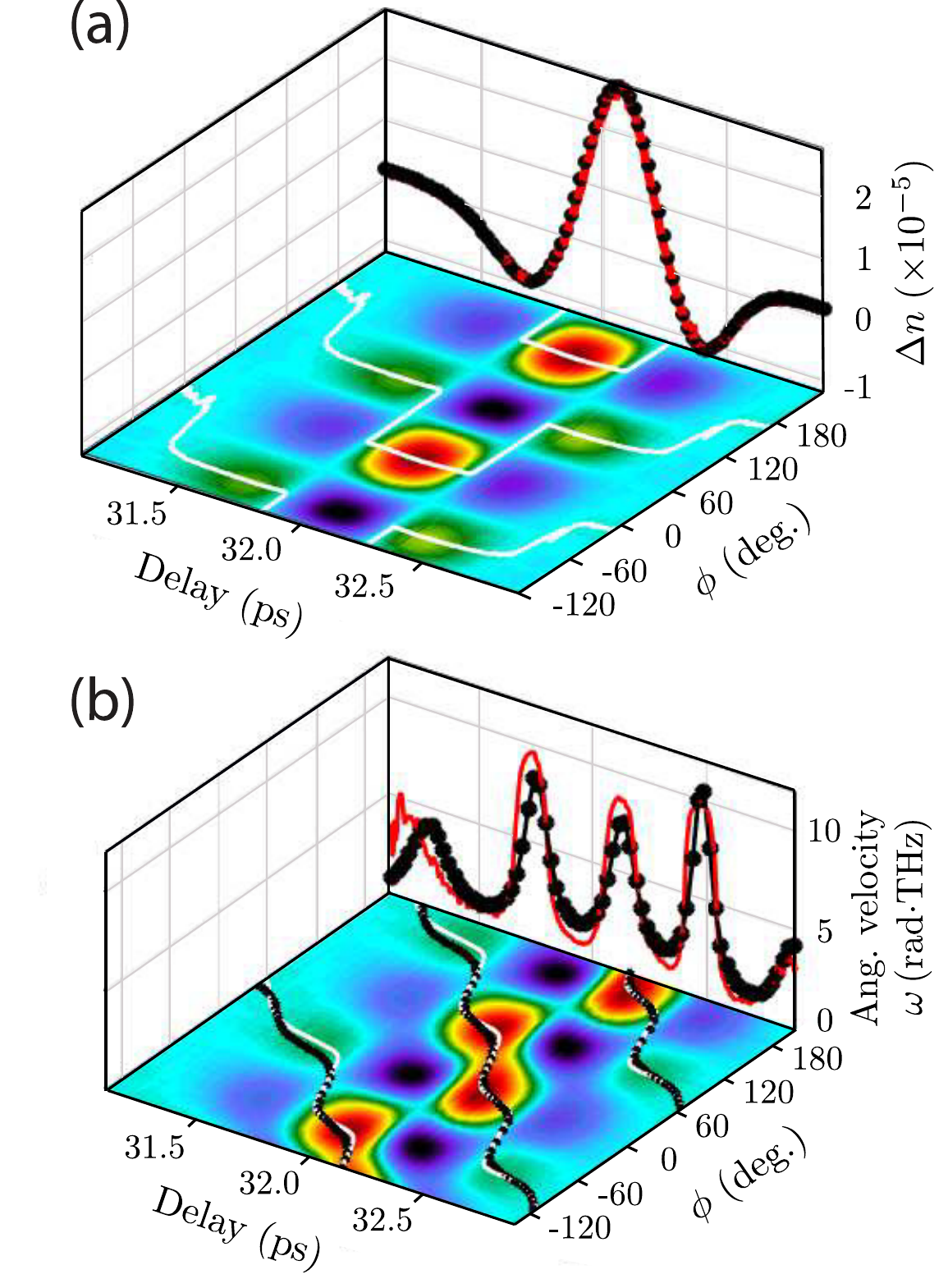}
  \caption{Temporal evolution of the angular distribution of the signal captured by the camera for a gas sample composed of a) aligned  and b) spinning molecules. The white solid lines and black squares refer respectively to the measured and calculated orientation angle $\phi$  of the 
  molecular axis with respect to the  $x$ axis.  a) Birefringence $\Delta n$ and b) measured (red solid line) and calculated (black filled circles) angular velocities $\omega=\partial_t\phi$  of the molecular axis  are depicted in the rear panels.}
  \label{fig-exp_distrib}
\end{figure}
 The birefringence and orientation of the molecular sample can be fully assessed by  analyzing the angular dependence of the images $M(R,\Phi)$, where $R$ and $\Phi$ are the radial and azimuthal coordinate of the camera, respectively. The angular dependence $S(\Phi)$ at each delay  $\tau$ is obtained from the corresponding image by summing the signal over $R$. Figure \ref{fig-exp_distrib}a [resp. \ref{fig-exp_distrib}b] depicts the delay-dependent angular signal  measured with a  pump pulse exhibiting a fixed  (resp. twisted) linear polarization. Assuming a cylindrically symmetric beam intensity pattern, one can show using the Jones matrix formalism that the angular dependence of the signal intensity $S(\Phi)$ can be ideally written
\begin{equation}
S(\Phi)\propto 1+\textrm{sin}\psi\ \textrm{sin}\left[2(\Phi-\phi)\right],
\label{EqJones}
\end{equation}
where $\psi$ is the phase difference experienced by the probe field components  along and perpendicular to the slow axis of the medium, i.e. the molecular axis, the latter making an angle $\phi$ with respect to the $x$ axis of the laboratory frame (see  Section \ref{sssec:spinning}).  It  is clear from Eq.\,\eqref{EqJones}  that all information about the molecular sample is embedded in the term oscillating at $2\Phi$. For instance,  we can see that the signal is maximum for $\Phi=\phi+\pi/4\mod{\pi}$, as described in section\,\ref{ssec:method}, and that  the amplitude of the oscillation is directly proportional to the dephasing $\psi$, which is related to the alignment factor  $\left<\cos^2\phi\right>$. For the data analysis, it is convenient  to cast  $S(\Phi)$  in the form
\begin{equation}
S(\Phi)\propto 1+\frac{\textrm{sin}\psi}{2}\textrm{e}^{-i\left[\pi/2+2\phi\right]}\textrm{e}^{2i\Phi}+\frac{\textrm{sin}\psi}{2}\textrm{e}^{+ i\left[\pi/2+2\phi\right]}\textrm{e}^{-2i\Phi},
\label{EqJones2}
\end{equation}
which is none other than the Fourier series of the signal  defined as
\begin{equation}
S(\Phi)=\sum_{n=-\infty}^\infty{S_n\textrm{e}^{-in\Phi}},
\end{equation}
where $S_n$ is the $n^{\textrm{th}}$ complex valued coefficient of the Fourier series. The information about the molecular sample ($\phi$ and $\textrm{sin}\psi$) can be directly retrieved from the second Fourier coefficient $S_2$
\begin{eqnarray}
\begin{aligned}
&|\textrm{sin}\psi|\propto|S_2|,& \\
&\phi=\frac{\textrm{arg}\left(S_2\right)}{2}-\pi/4\,\mod{\pi}.&\\
\end{aligned}
\label{Analysis}
\end{eqnarray}
Note that Eqs.\,\eqref{Analysis}  can also be used as a reasonably  good approximation of  $\sin\psi$ and $\phi$ in case   the angular dependence of the experimental signal is affected by any imperfection of the detection setup, as for instance a residual ellipticity of the probe beam 
or a non perfect  cylindrical laser beam (see Sec. S2.C, Supporting Information). For completing the analysis of the experimental data,  we have also estimated with the help of   Eqs.\,\eqref{Analysis} the  birefringence  $\Delta n=\frac{\rho}{2\epsilon_0}\Delta \alpha^\prime$ ($\rho$ is the number density,  $\epsilon_0$ is the vacuum permittivity, and $\Delta \alpha^\prime$ is the  difference of polarizability  calculated between the neutral axes of the medium) and the  instantaneous angular velocity  of the  medium  $\omega(t)=\partial_t\phi$. $\Delta n$ and $\omega(t)$ are depicted in Figures \ref{fig-exp_distrib}a and 6b, respectively.

\begin{figure}
  \includegraphics*[width=\linewidth]{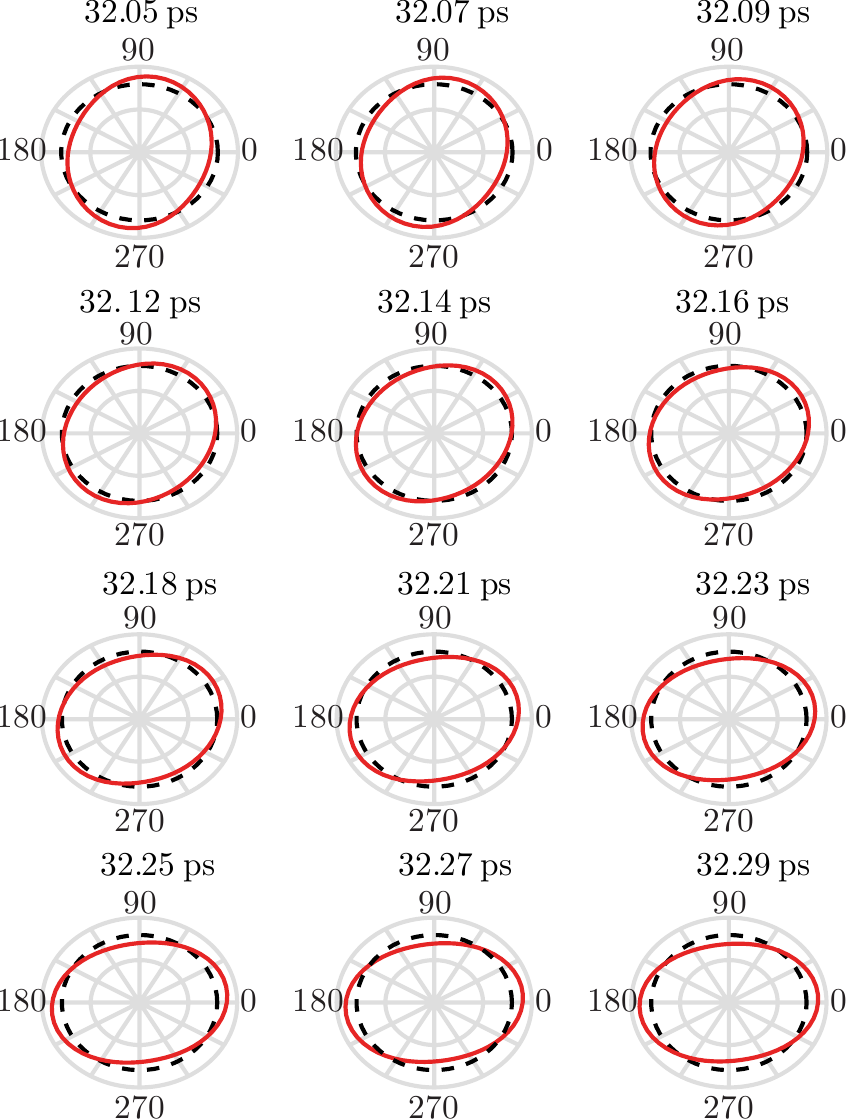}
  \caption{Angular distributions $P(\phi,t)$ of room-temperature molecules (solid red lines) depicted in the polarization plane of the pump field with twisted polarization at different delays. The pictures correspond to $3T_{\textrm{r}}/4$ revival of CO$_2$ for the delays and experimental conditions corresponding to Figure \ref{fig-udr}. The black dashed lines are used to guide the eye and represent the isotropic distribution.}
  \label{fig-theo_distrib}
  \end{figure}

 In order to test the accuracy of the optical imaging technique, we have computed the time-dependent angular distribution of the molecules, from which the birefringence $\Delta n$   of the medium and  the  angular velocity $\omega$ of the molecular axes can be estimated. As shown in Figure \ref{fig-exp_distrib},  the calculated values are in very good agreement with the data extracted by analysing the recorded images. The latter are also well reproduced by the theory if one compares the records of  Figure \ref{fig-udr} with the simulations of Figure \ref{fig-theo_distrib} representing  the angular distributions of the spinning molecules  calculated at the same delays. 

\section{Conclusion}
In summary, we have developed an optical method for imaging  quantum rotors  that are angularly localized in space. To the best of our knowledge, it is the only all-optical imaging technique reported so far enabling the visualization of ultrafast molecular rotation.  It relies on  the spatially resolved detection of transient birefringence performed in a pump-probe configuration.  The key element of the detection is  a  vortex half-wave  retarders plate that  combined with a linear polarizer and a camera allows to  capture the  orientation of the molecules across the beam of a femtosecond laser pulse. To this aim, the latter is  circularly polarized before its interaction with the molecular  gas.  The imager is successfully employed to locate the orientation of the molecules after they have been either aligned or spun by an extra short laser pulse. By tuning the delay between the two laser pulses, one can record, on a single shot basis, the ultrafast coherent rotation of  a molecule ensemble.  Besides the orientation of the medium, it is  demonstrated that useful  additional information,  as the birefringence  or average angular velocity of the aligned molecules in the medium, can be extracted from analyzing the recorded images. More generally, the present optical method  locates  the neutral axes of a birefringent medium and therefore is in principle applicable  to any non isotropic transparent medium including  molecular superrotors\cite{Korobenko2014}, nanostructured\cite{Hakobyan8_2014},  macroscopic objects and metamaterials\cite{Dogariu24_2012}.

\label{sec:conclusion}

\bibliographystyle{lpr}

\end{document}